\documentclass[12pt]{iopart}

\usepackage{graphicx}  
\begin{document}

\title{Group Chase and Escape}

\author{Atsushi Kamimura$^1$ and Toru Ohira$^{1,2}$}

\address{$^1$Department of Applied Physics, The University of Tokyo, 
7-3-1, Hongo, Bunkyo-ku, Tokyo, 113-8656, JAPAN}
\address{$^2$Sony Computer Science Laboratories, Inc., 3-14-13, Higashi-gotanda, Shinagawa-ku,
Tokyo, 141-0022, JAPAN}
\ead{kamimura@serow.t.u-tokyo.ac.jp}
\begin{abstract}
We describe here a new concept of one group chasing another, called ``group chase and escape", 
by presenting a simple model. We will show that even a simple model can demonstrate
rather rich and complex behavior. 
In particular, there are cases in which an optimal 
number of chasers exists for a given number of escapees (or targets) 
to minimize the cost of catching all targets. 
We have also found an indication of self-organized
spatial structures formed by both groups.

\end{abstract}

\maketitle

\section{Introduction}

Issues relating to chase and escape have a long history and 
have often led to interesting mathematical results \cite{isaacs65,basar99,nahin07}. 
On this theme, one chaser pursues a single target, such
as a war-ship chasing a pirate vessel. 
Various cases have been studied with different
boundaries and conditions, speed ratios, and so on. Often they have led to
rather complex trajectories and posed challenging mathematical problems for 
obtaining analytical expressions. 
This topic has developed by being driven by mathematical interests 
together with its various applications, such as military ones. 
Developing computational systems
and techniques enables one to deal with quite complex situations for the one-to-one case.
The cases in which there are multiple chasers and/or escapees are much less studied.
There are predator-prey models that consider multiple entities chasing a single target
or prey \cite{krapivsky96,oshanin09}. Also, 
there are some recent studies on to several evaders, 
which have been done 
in the context of game theory, robotics, and multi-agent systems \cite{hespanha99,vidal02}. 
From a point of view of physical systems, there is a study of collective motion of Brownian
particles with pursuit-and-escape interactions \cite{romanczuk09}. However, in this study 
particles are not grouped separately as chasers and escapees, and to the authors' knowledge, such
models have not been investigated.

Against this background, the main theme of this letter is to propose a paradigm of research
problems associated with one group chasing another, which we term ``group chase and escape".

From the viewpoint of physical systems, the problem of group chase and escape is an extension of 
studies of granular materials \cite{jaeger96,degennes99,kadanoff99} to traffic problems \cite{wolf96,sugiyama08}, which have been actively 
investigated in recent decades. In the traffic problems, so-called ``self-driven particles''
are the basic constituents rather than physical particles like in granular materials.
In the field of statistical physics, a focus of interest is investigating
how the microscopic dynamics produces macroscopic behavior such as phase transitions, 
scaling, and pattern formations.
We are extending each unit further by giving them the aim of chasing or escaping.

\section{Model}

Let us describe our model.
We consider a two-dimensional square lattice $L_x \times L_y$ with periodic boundary conditions.
Each site is empty or occupied by one particle: a chaser or escapee (target).

\begin{figure}
\begin{center}
\includegraphics[height=5cm]{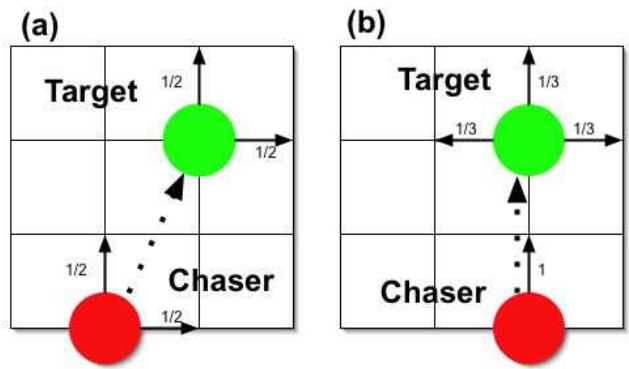}
\caption{Hopping rules for chasers and targets. While chasers hop to close in on their nearest targets,
targets hop to evade their nearest chasers. 
Dotted arrows from chaser to target indicate that chaser hops to close in on target. 
Solid arrows show possible hopping directions with indicated probabilities.
(a) Generally, they have two choices.
(b) When chasers or targets are in same $x$ or $y$-axis, chasers have one choice, while targets have three choices. }
\label{fig:1}
\end{center}
\end{figure}

The chasers and targets play tag by hopping between sites 
in accordance with the following rules.
Every target wants to evade its nearest chaser.
Let us denote the positions of target and chaser as $(x_T, y_T)$ and $(x_C, y_C)$, respectively.
For each target, the distance to each chaser is calculated as $d = \sqrt{(x_T - x_C)^2 + (y_T - y_C)^2}$.
Then a chaser in a minimum $d$ is the nearest to the target.
Here, if there are more than two chasers equally near (-equal $d$-), 
we choose one of them randomly.
Then the target hops to its nearest site in the direction that increases the distance from the chaser. 
The hopping rule is shown in Fig. \ref{fig:1}.
Generally, the target has two possible sites to which to hop, as shown in Fig. \ref{fig:1}(a).
In this case, one of the two sites is chosen with an equal probability $1/2$. 
If the $|x_T - x_C|$$(|y_T - y_C|)$ is zero, then the target has three possible sites to increase the distance,
 one of which is chosen with an equal probability $1/3$ (-see Fig. \ref{fig:1}(b)-).

The rule for chasers is that they hop to close in on their nearest targets.
In the same manner as the targets, we determine the nearest target for every chaser.
The chaser hops to its nearest site that decreases the distance.
Generally, the chasers choose one of two possible sites to which to hop with an equal probability $1/2$.
Here, if the $|x_T - x_C|$$(|y_T - y_C|)$ is zero, then the chaser hops only in the $y-$$(x-)$ direction 
because hopping in the $x-$$(y-)$ direction increases the distance.

Except in catch events explained below, chasers and targets cannot hop to the nearest sites if the sites are
occupied, so they remain in their original sites. 
Here, we first choose one of the nearest sites by the probabilities above, 
so that the particles do not move even if the other nearest sites are empty.

When a target is in a site nearest to a chaser, the chaser catches the target by hopping to the site, and then the target is removed from the system. 
After the catch, the chaser pursues the remaining targets in the same manner.

In accordance with the above hopping step, every chaser and target hops by one site. 
In the simulations, we first determine the next hopping site for chasers and targets.
Then we move chasers. Here, if a chaser hops to a site a target occupies, 
the chaser catches the target. 
After this, we move targets. 
Here, the update is done in a random sequential order. 
Initially, $N_C^0$ chasers and $N_T^0$ targets are randomly distributed in the lattice.
While the number of chasers, $N_C$, remains a constant $N_C^0$, the number of targets, $N_T$, monotonically decreases along with the catches. 
Simulations are carried out until all targets have been caught by chasers, i.e., $N_T = 0$.
The results are averaged over $10^4$ runs.

In addition to the chase-and-escape hopping model, we consider random walk processes in which a particle hops to one of the four nearest sites with an equal probability $1/4$, irrespective of the positions of chasers and targets.
In this model, targets are caught when a chaser in the nearest site tries to hop to the site of the target. 
As we will explain below, we examine different cases: both chasers and targets or either of them follows the random walk.

Before examining the chase-and-escape processes, let us note the diffusion model 
in which both chasers and targets follow the random walk processes.
In this model, dynamics of the number of targets $N_T$ can be interpreted as a reaction-diffusion system in which
targets are annihilated when they meet chasers, leading to a rate equation,
\begin{equation}
\frac{dN_T}{dt} = -kN_TN_C,  
\end{equation}
where $k$ denotes a rate constant. As the number of chasers remains a constant, 
the solution gives $N_T(t) = N_T(0) \exp(-kN_C t)$. 
By rewriting the equation as 
$d\log(N_T)/dt = - k N_C$, we can also note the effect of fluctuations.
If we assume that the rate constant $k$ fluctuates with a normal distribution, 
the fluctuation of $N_T$ will be given by a log-normal distribution.

\section{Simulation Results}

\subsection{Cost of group chase} 

\begin{figure}
\begin{center}
\rotatebox{270}{
\includegraphics[height=8cm, clip]{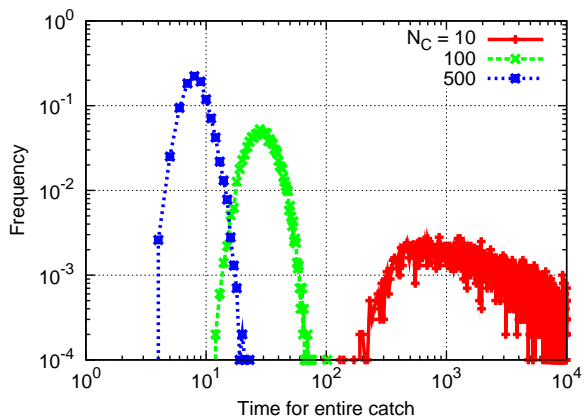}}
\caption{Distribution of time for entire catch $T$ for different $N_C$. 
Parameters are $L_x = L_y = 100$, $N^0_T = 10$.}
\label{fig:2}
\end{center}
\end{figure}

To characterize the nature of our model, first let us 
look at the time length for the entire catch $T$. In other words,
$T$ is the time it takes the chasers to catch all the targets.
Their distribution is shown in Fig \ref{fig:2}. 
Since $T$ can also be interpreted as a ``lifetime" of the final target, Fig. \ref{fig:2} gives the
probability distribution of its lifetime. When $N_C$ is larger than $N_T$, we note this distribution 
basically shows a parabolic shape in the log-log scales, suggesting a log-normal distribution. 
This distribution is deduced even in a pure diffusion case noted above.
However, as $N_C \approx N_T$, it deviates from the log-normal distribution, reflecting the effect of chase and escape.
 
\begin{figure}
\begin{center}
\includegraphics[width=14cm, clip]{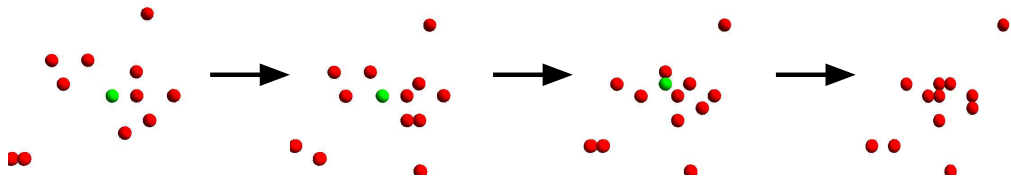}
\caption{Snapshots of catch event with time evolution from left to right. 
Red and green circles denote chasers and target, respectively.}
\label{fig:3}
\end{center}
\end{figure}

The average length of time decreases as the number of chasers increases.
Here, the speeds of chasers and targets are equal, $V_C = V_T$.
In this case, an individual chaser cannot catch up with targets, 
so it cannot finish the job by itself.
Instead, a group of chasers catches a target by surrounding it so that the target cannot escape from them.
This typical catch event is shown in Fig. \ref{fig:3}. 
Although an individual chaser independently tries to catch a target, 
it appears as if the group of chasers cooperates to catch a target.

\begin{figure}
\begin{center}
\rotatebox{270}{
\includegraphics[height=8cm, clip]{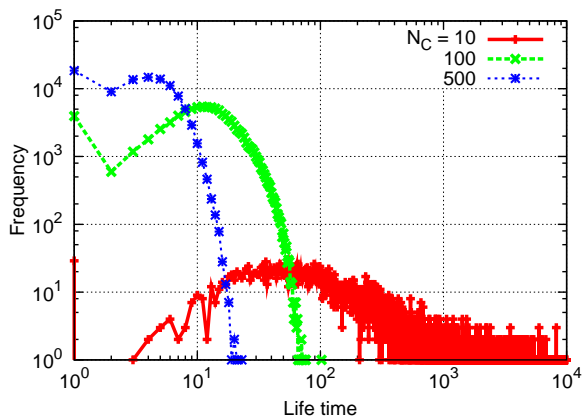}}
\caption{Lifetime distribution of targets. Parameters are identical with Fig. \ref{fig:2}}
\label{fig:4}
\end{center}
\end{figure}

If we look at the lifetime distribution of all targets, then we obtain
the results in Fig. \ref{fig:4}. 
The distribution first shows large drops at the left, 
then increases, and peaks at a typical time. After the peak, 
it decreases again.    
The first drop suggests a large number of targets, 
the lifetime of which is one. 
This is because in the initial condition, targets can be positioned in the sites
nearest to chasers, i.e., $d=1$.
Thus the targets are caught by the chasers in the next step, and
such an unlucky target's lifetime equals one.
If the initial distances between targets and chasers are larger than two, 
the targets can momentarily evade the chaser, causing the drop.
After the drop, the distribution increases, and 
we can see this distribution peak.
The value of these peak positions can be inferred as a typical lifetime.
This lifetime represents a timescale in which the group of chasers gathers 
around targets from the initial conditions and catches them.
As the number of chasers increases, the timescale decreases.  
For comparison, we note the lifetime distribution in other cases.
If we look at the distribution in the random walk model, 
it decreases exponentially, so it does not drop or peak.
If we examine a model in which chasers close in on targets but targets follow
the random walk processes, the peak appears to represent the timescale.
However, the drop does not appear because the targets do not evade chasers, 
so there is no drastic drop from the lifetime of one to two. 

\begin{figure}[htbp]
\begin{tabular}{cc}
\begin{minipage}{0.5\hsize}
\begin{center}
\rotatebox{270}{
\includegraphics[height=8cm]{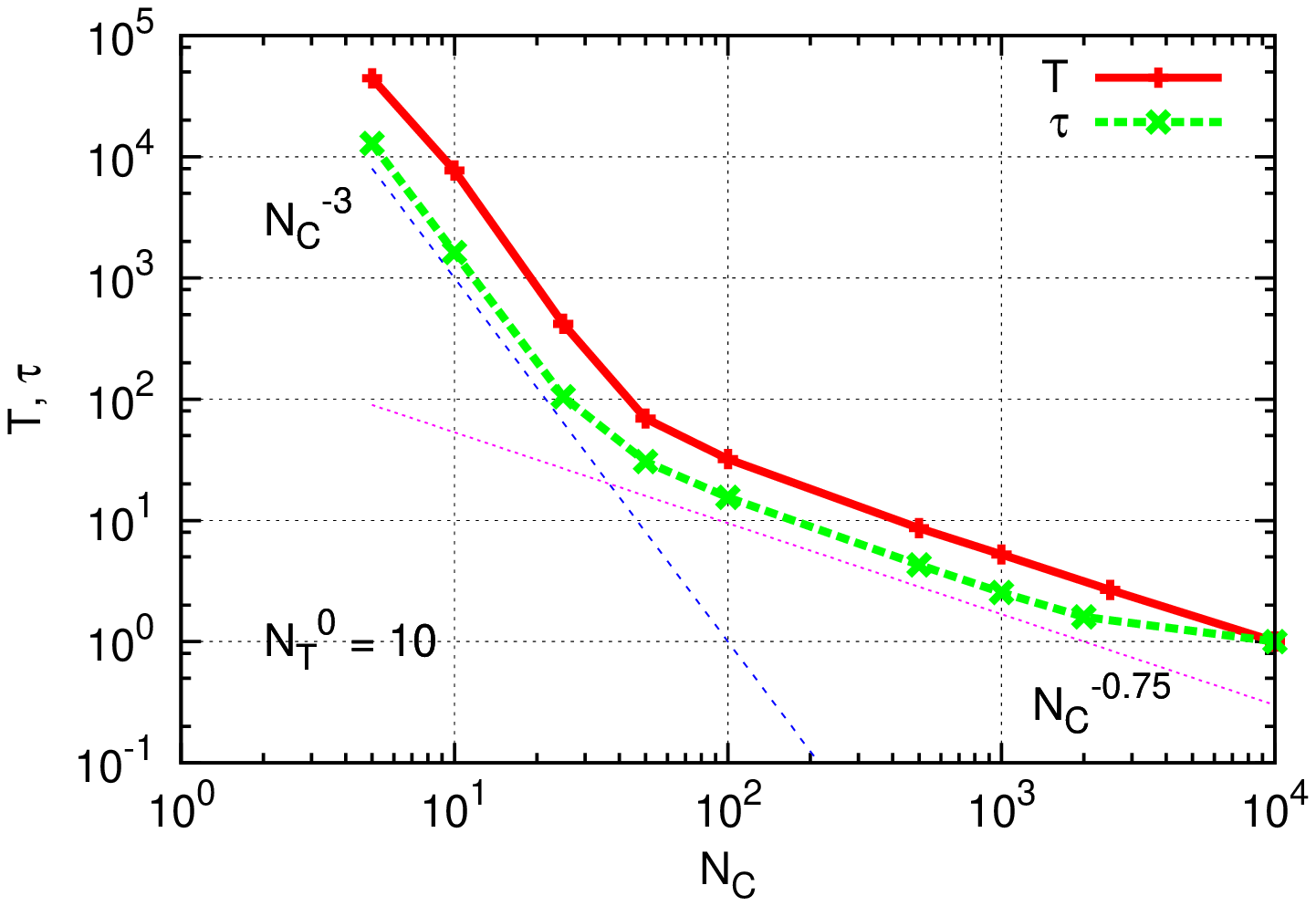}}
\caption{Lifetime of final ($T$) and typical ($\tau$) targets for fixed $N_T=10$.}
\label{fig:4a}
\end{center}
\end{minipage}
\begin{minipage}{0.5\hsize}
\begin{center}
\rotatebox{270}{
\includegraphics[height=8cm]{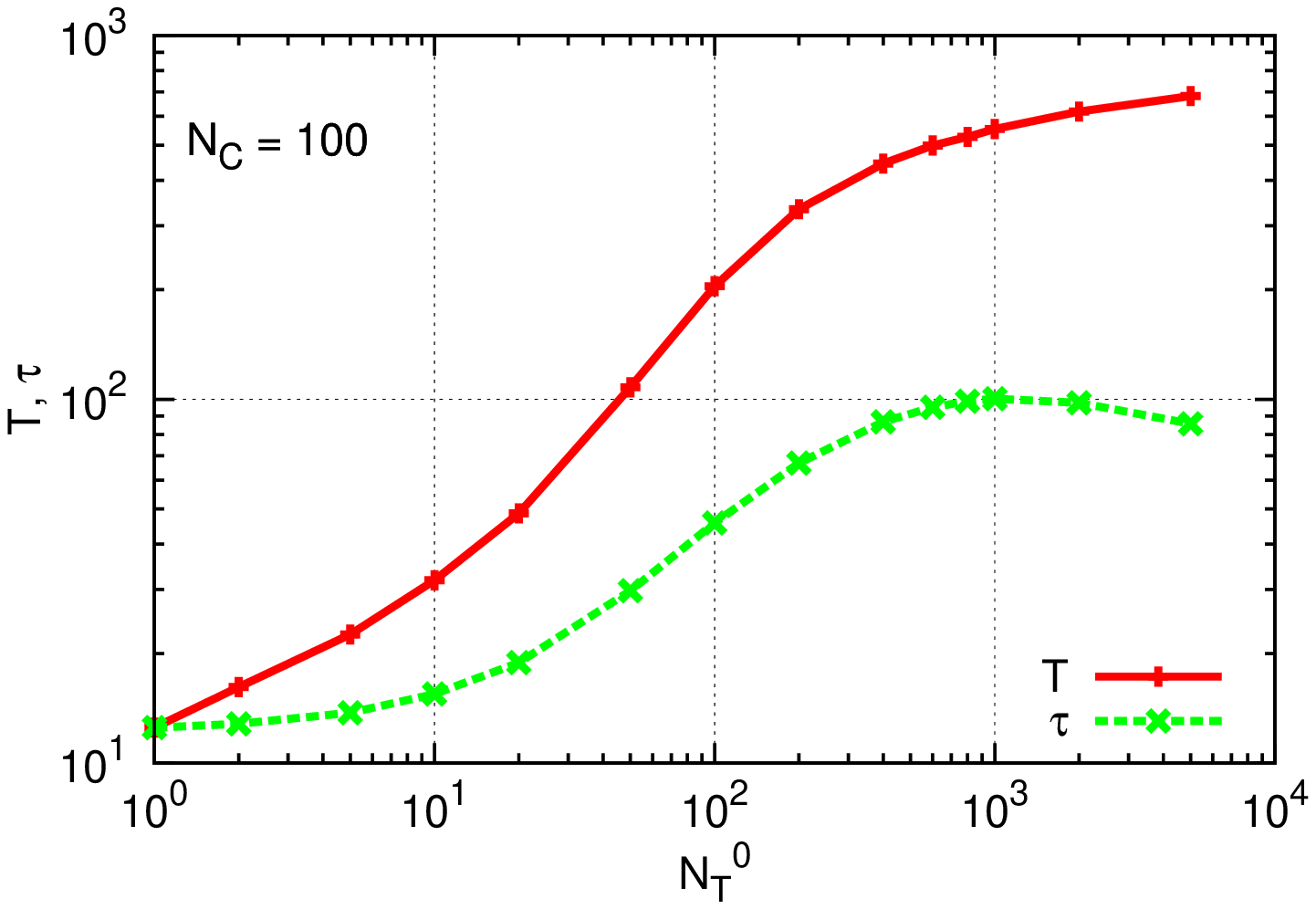}}
\caption{Lifetime of final ($T$) and typical ($\tau$) targets for fixed $N_C=100$.}
\label{fig:4b}
\end{center}
\end{minipage}
\end{tabular}
\end{figure}

We investigate how the lifetimes of the final and typical targets 
change with $N_C$ and $N_T$. 
The lifetime of the final targets is $T$. 
The typical lifetime is defined as $\tau_t = \Sigma t ( N_T^t - N_T^{t-1} )/N_T^0$, where 
$N_T^t$ denotes the number of targets at $t$ so that $( N_T^t - N_T^{t-1} )$ represents
the number of targets of lifetime $t$.
In Fig. \ref{fig:4a}, we show how $T$ and $\tau$ change with
$N_C$ for a fixed $N_T^0=10$. 
Both $T$ and $\tau$ behave similarly. 
For moderately small $N_C$, the lifetimes decrease as
$N_C^{-3}$.
This regime is where $N_C$ is of the order of ten times larger than $N_T^0$.
We note here that in the previous study \cite{oshanin09} of one target escaping from a group of chasers, 
survival probability scales with cube of the number of chasers.
However, as $N_C$ increases further, 
the lifetimes show crossover to slower decreases, which are approximately fitted by $N_C^{-0.75}$. 
In the right end, both of them approach one where the sites are filled with chasers so that
both typical and final targets can survive only by one time step.

\begin{figure}
\begin{center}
\rotatebox{90}{
\includegraphics[height=15cm, clip]{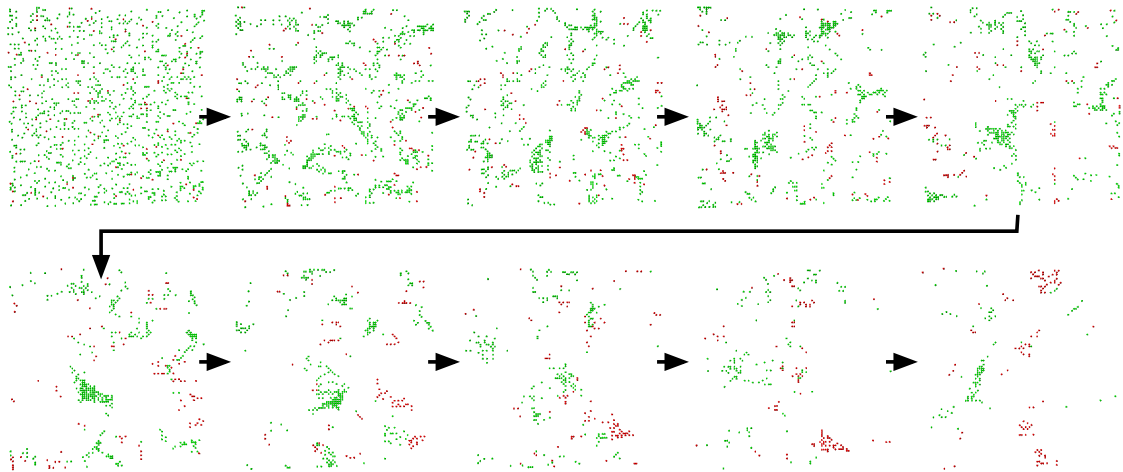}}
\caption{Time evolution of system for $N_C = 100$ and $N_T^0 = 1000$.
Red and green points represent chasers and targets, respectively.
Time evolves from initial condition in left top to right, left bottom to right bottom.}
\label{fig:4c}
\end{center}
\end{figure}

In Fig. \ref{fig:4b}, we show how the lifetimes change with $N_T^0$ for a fixed $N_C=100$. 
As $N_T$ increases, the lifetime of final targets monotonically increases.
However, the lifetime of typical targets peaks around $N_T = 10^3$ and slightly 
decreases again.
Around this peak, we show typical snapshots in Fig. \ref{fig:4c}. 
From the initial condition in the left top, 
targets evade chasers at first by producing clusters of groups of targets. 
From the second snapshot, 
we can see the clusters of targets appear where targets aggregate.
Then a group of chasers gets closer to the clusters, catching targets.
It is intuitively efficient for the group of chasers 
to catch targets 
by surrounding the cluster of targets because a number of targets can be caught 
by once.
The peak of the lifetime may represent such effect.

Also, it is of interest to know the ``right" number of chasers $N_C$
for a given number $N_T^0$ of targets. We have evaluated this
by focusing on the quantity $c = N_C T / N_T^0$. This quantity represents
the unit cost for the group of chasers to finish the job per target. (The 
amount of work-hours $N_C$ for which chasers are deployed (total cost) divided by the number
of targets $N_T$ .)

\begin{figure}
\begin{center}
\rotatebox{270}{
\includegraphics[height=8cm, clip]{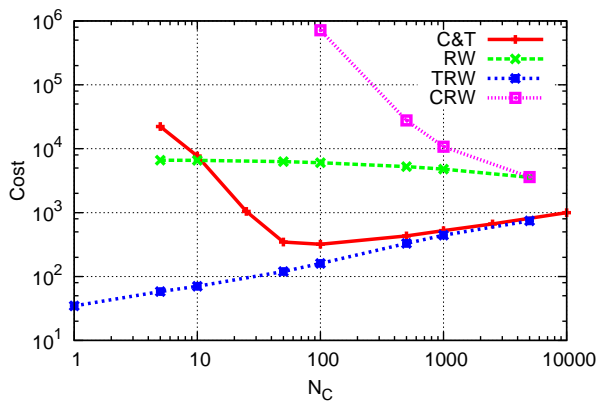}}
\caption{Cost $c$ vs. number of chasers for $N_T^0 = 10$ in following four cases: original chasers and targets (C\&T), both are random walkers (RW), targets are random walkers (TRW), and chasers are random walkers (CRW). }
\label{fig:6}
\end{center}
\end{figure}

We have plotted this unit cost function for different cases. 
In Fig. \ref{fig:6}, we examine the cost by changing $N_C$ for a fixed $N_T^0$.
When we see the original chase and target case (C\&T),
there is a minimum in this unit cost. This means there is an optimal number
of chasers $N_C^*$ to finish the given group chase task most efficiently.
When the targets are as fast as the chasers,
an individual chaser cannot catch up with targets, so it cannot finish the job by itself.
Instead, a group of chasers catch a target by surrounding it so that the target can not escape from them. 
In this case, having sufficiently more chasers than targets is necessary to finish the job efficiently.
On the other hand, as the number of chasers exceeds the optimal number necessary to surround the targets,
excessive chasers result in the cost increasing.  
The right side of the figure $N_C=9990$ confirms that 
the system is fully occupied, so 
the targets are caught in one simulation step, leading to the cost $c = 1*9990/10 = 999 \sim 10^3$. 

Such a minimal cost is realized as a result of both chase-and-escape processes.
In Fig. \ref{fig:6}, we also show the costs in different cases: both or either chaser and target follow 
a random walk process.
We see that when the targets are random walkers (TRW), the cost monotonically rises along with 
number of chasers. On the other hand, when the chasers (CRW) or both (RW) are random walkers, 
they monotonically decrease.

\subsection{Issues of range of each chaser}

\begin{figure}
\begin{center}
\rotatebox{270}{
\includegraphics[height=8cm, clip]{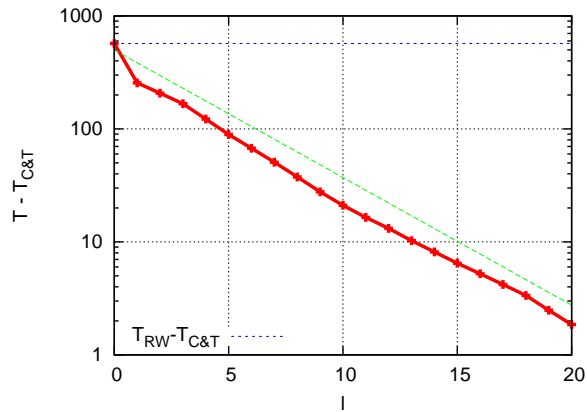}}
\caption{$T-T_{C\&T}$ as function of search range $l$. Parameters are $N_C=100$ and $N_T^0=10$.}
\label{fig:R}
\end{center}
\end{figure}

We can extend our model to include the search range of each chaser.
In the current model, chasers can find targets over an unlimited distance. 
However, in reality, chasers search for targets in their vicinities.
This is also the same for targets. Targets can recognize the existence of nearby chasers.
We introduce the search range $l$ as follows.
When a chaser searches for the nearest target, the search 
area is limited to the range $\sqrt{(x_T-x_C)^2+(y_T-y_C)^2} < l$, where 
$x_i, y_i$ denote the positions of targets ($i=T$) and chasers ($i=C$) in $x$ and $y$-directions, respectively.
If the chaser finds a target in the search range, it moves with
the chase-and-escape hopping. If not, it follows the random-walk hopping.
For the movement of targets, the search range can be introduced in the same manner.
If the value of $l$ equals zero, the movement is equivalent to the random walkers. 
On the other hand, the movement approaches to the previous model as the range increases to the system size. 
In Fig. \ref{fig:R}, we show $T - T_{C\&T}$ as a function of $l$, where 
$T_{C\&T}$ is the time for entire catch without search range limits. 
When $l=0$, the time is equal to that of random walkers.
As $l$ increases, the difference decreases exponentially approaching zero.

\begin{figure}
\begin{center}
\includegraphics[height=5cm, clip]{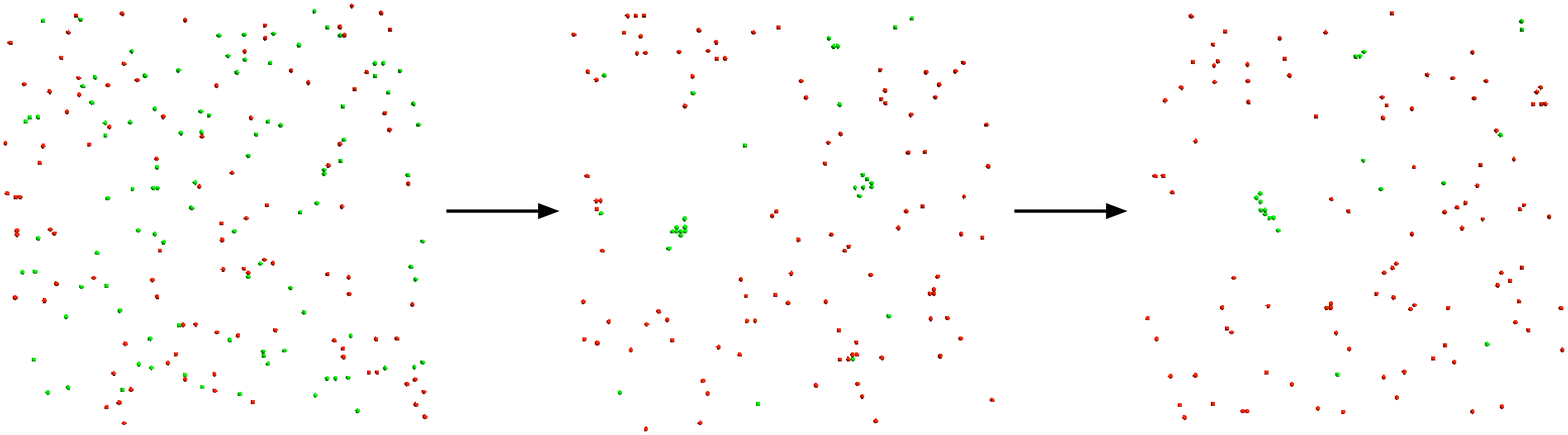}
\caption{Snapshots of system with time evolution from left to right. While targets have unlimited search range, 
chasers have the search range $l = 5$.  The numbers of chasers and targets are fixed to $N_C = N_T^0 = 100$.}
\label{fig:7}
\end{center}
\end{figure}

When the search range is different between chasers and targets, 
the systems can exhibit interesting behavior. 
Figure \ref{fig:7} shows one such example. 
Here, we assume an unlimited search range for targets, while 
the range for chasers is sufficiently short.
For an appropriately low number of chasers, targets gather in relatively low-density areas of chasers and 
momentarily hide from chasers because the short-range chasers cannot recognize their existence.
After a long time, chasers can find the group of targets and finally catch them.
Examining the catching processes in relation to such formations of spatial patterns can be an interesting topic. 

\subsection{Issues of long-range chaser doping}

\begin{figure}
\begin{center}
\rotatebox{270}{
\includegraphics[height=8cm, clip]{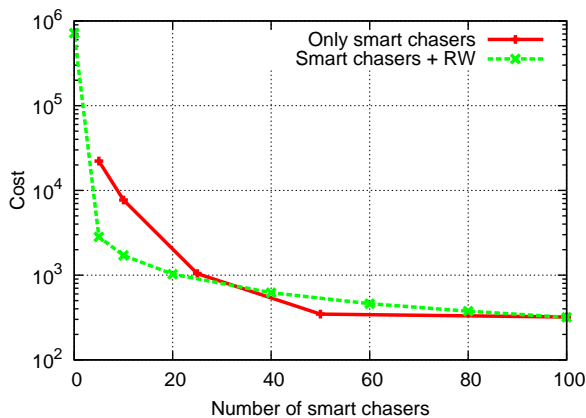}}
\caption{Cost $c = TN_C/N_T^0$ by changing number of smart chasers $N_C$ for fixed $N_T^0 = 10$.}
\label{fig:8}
\end{center}
\end{figure}

We also consider another extension of our model.
This extension is to distribute the search ability among chasers.
For example, in the group of chasers, 
some have a long search range, while the others follow the random-walk hopping or have 
a short search range.
We look at the cost in such an example.
The group of chasers consists of two types: smart chasers and random walkers.
The smart chasers have an unlimited search range.
On the other hand, the random walkers have search ranges of zero.
In Fig. \ref{fig:8}, we show the cost by changing the number of smart chasers $N_C^s$ with
a fixed $N_C$ so that $N_C - N_C^s$ random-walking chasers also join the catch in the system.  
For comparison, we also show the case in which only $N_C^s$ smart chasers are in the system 
and play tag.
Here we assume that targets have unlimited range.
The left end $0$ corresponds to the case in which all of the chasers are random walkers.
The right end $100 = N_C$ corresponds to the case in which all chasers have limitless ranges.
As the ratio of smart chasers increases, the cost monotonically decreases.
Interestingly, even a small number of smart chasers, say five to ten, 
drastically drop the cost. 
Comparing the two cases is also interesting. 
If only a small number of smart chasers are available, which strategy will be better: 
let only the smart ones chase targets, or have random walkers join them?
The group of random walkers also contributes to the catch events so 
we initially expect that the latter case is better. 
However, if we have to pay the salary per working hour to the chasers, 
the more chasers join, the more we have to pay. 
From the viewpoint of cost, we can say the latter case is more efficient.
However, as the number of smart chasers increases to $30$ or $40$, the former strategy is better.

\subsection{Issues of hopping fluctuations}

In the previous subsections, we introduced the search range of chasers and targets.
The search range can control the model departing from the random walk 
to the chase-and-escape models.
We propose here another extension to introduce such a parameter.

In the original hopping rule, chasers/targets must choose the next site
to decrease/increase distances to the nearest targets/chasers.
We introduce fluctuations to these decisions.
When a chaser chooses the next hopping site, 
the hopping probabilities are defined as follows.
For each of the four nearest-neighbor sites, we define $\Delta l_i = \pm 1$ where $i$ denotes the indexes of 
the four sites$(i=1,2,3,4)$.
If hopping to a site $i$ decreases the distance to the nearest target, 
we assume $\Delta l_i = -1$. If it increases, we assume $\Delta l_i = 1$.
Then we define the hopping probability of the chasers 
as $p^C_i = \exp(-\Delta l_i/T_f)/\Sigma_i \exp(-\Delta l_i/T_f)$,
where we introduce $T_f$ as a ``temperature".
In the same manner, we define $\Delta l_i$ for targets.
When hopping to a site $i$ increases the distance from the nearest chaser, 
we assume $\Delta l_i = 1$. If it decreases, we assume $\Delta l_i = -1$.
We define the hopping probability for targets as 
$p^T_i = \exp(\Delta l_i/T_f)/\Sigma_i \exp(\Delta l_i/T_f)$.

When the temperature $T_f$ is sufficiently high, 
the value of $\Delta l_i$ is not relevant and the hopping probability 
is approximately equal to the random-walk model $1/4$.
As the temperature decreases, the hopping probability increases for chasers/targets 
to decrease/increase the distance, approaching the chase-and-escape model.

\begin{figure}
\begin{center}
\rotatebox{270}{
\includegraphics[height=8cm, clip]{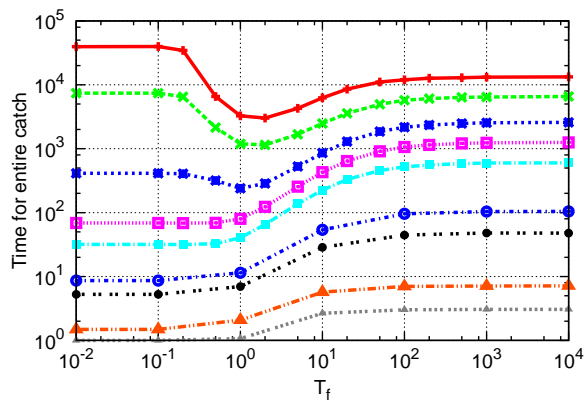}}
\caption{Time for entire catch as function of temperature $T_f$. 
Lines from the above to bottom are for $N_C = 5, 10, 25, 50, 100, 500, 1000, 5000, 9990$.
For all lines, we fix $N_T^0 = 10$.}
\label{fig:T}
\end{center}
\end{figure}

Figure \ref{fig:T} shows the time for entire catch as a function of the temperature for 
different values of $N_C$. The value of $N_T^0$ is fixed to 10.
For all lines, in the left and right ends ($T_f = 10^{-2}$ and $10^4$), 
the values of time for an entire catch are equal to those of the original chase-and-escape and random-walk models, respectively.
In between, we found interesting behavior. 
When $N_C$ is large(-for lower lines-), the time monotonically increases from left to right.
However, when $N_C$ becomes small to the order of $N_T^0$ (-for upper lines-), they show minimum around $T_f = 1$.
Here, we note that shapes of the distributions also change with $T_f$.
But we confirmed that the distribution with $T_f \sim 1 $ is clearly located at smaller value of time
 compared to those of the chase-and-escape and random-walk models.

Interestingly, a certain amount of fluctuations reduces the time, making it easier for chasers to catch targets. 
We may relate this observation to a phenomenon called ''Stochastic Resonance'' \cite{wiesenfeld-moss95,bulsara96,gammaitoni98}. 
Stochastic resonance has been studied in various fields from the stance that 
an appropriate level of noise or fluctuations can provide constructive or 
beneficial effects. In particular, we note the similarity of collective effects 
of stochastic resonance with a simple model of computer network traffic, 
where the appropriate level of fluctuations in the direction of passing 
packets by routers led to reducing the overall congestion of the network\cite{ohira-sawatari98}

\section{Discussion}

In this paper, we introduced a new concept termed ``group chase and escape" by
presenting a simple model. 
By developing the cost function, we found characteristic behavior of group chase processes 
and evaluated efficiencies that have an optimum number of chasers.
The values of the function were also compared among several cases, including the random walk. 
Our results confirm that the microscopic chase-and-escape rule
has a scheme completely different from that of a reaction-diffusion system
and also is a promising new example of a ``self-driven particle" system.

We could consider various extensions of our model.

One interesting extension would be to include the effects of swarms \cite{bonabeau99,couzin05}.
By comparing cases in which the targets/chasers are together or solitary, 
efficient survival/catch-up strategies could be developed.
The advantages and disadvantages of forming swarms are commonly
studied in the fields of sociobiology, such as risk dilution.
We could also examine the role differentiations and cooperative behavior 
in such groups. Interesting applications can be considered to 
the behavior of army ants, which cooperate to hunt, form bridges, and so on. 

Another extension we can make would be to incorporate more complex strategies 
for the chases and escapes. Instead of sensing the nearest target/chaser, 
each side could use ``center of mass" of the locations, 
or more information about distributions of the opponent. 
Changing strategies on the basis of situations, 
such as the number of non-captured, could be also considered. 

We could also include the effects of information transmission delay 
for chasers and targets to grasp the other particles' positions. 
Delays often introduce unexpectedly 
complex effects into otherwise simple dynamical systems\cite{glass88,ohira00,balachandran09}.
Some examples of applications are the modeling of blood cell reproduction \cite{mackey77}, human
posture and balance controls \cite{ohira95,milton09a,milton09b}, traffic jams \cite{bando95,nakanishi97}, and so on. 
Delays could produce interesting behavior in the context of chases and
escapes.

\section*{Acknowledgement}
The authors would like to thank the members of Prof. Nobuyasu Ito's group at the University of Tokyo 
for their fruitful discussions. Thanks also goes to Prof. John G. Milton of the Claremont 
Colleges for introducing us to the work of Nahin \cite{nahin07}.


\section*{References}
\begin{thebibliography}{10}
\bibitem{isaacs65} Isaacs R {\it Differential Games} 1965 
(New York: Wiley)

\bibitem{basar99} Basar T and Olsder G {\it Dynamic Noncooperative Game Theory} 1999 
(Philadelphia: SIAM)

\bibitem{nahin07} Nahin P J {\it Chases and Escapes: The mathematics of pursuit and evasion} 2007 
(Princeton: Princeton University Press)

\bibitem{krapivsky96} 
Krapivsky P L and Redner S 1996
{\it J. Phys. A: Math Gen.}{\bf 29} 5347

\bibitem{oshanin09}
Oshanin G, Vasilyev O, Krapivsky P L and Klafter J 2009
{\it Proc. Nat. Acad. Sci} {\bf 106} 13696

\bibitem{hespanha99}
Hespanha J P, Kim H J and Sastry S 1999
{\it in Proc. 38th Conference on Decision and Control} 2432

\bibitem{vidal02}
Vidal R, Shakernia O, Kim J H, Shim D H and Sastry S 2002
{\it IEEE Trans. Robotics and Automation} {\bf 18} 662

\bibitem{romanczuk09}
Romanczuk P, Couzin I D and Schimansky-Geier L 2009
{\it Phys. Rev. Lett.}{\bf 102} 010602


\bibitem{jaeger96}
Jaeger H M, Nagel S R and Behringer R P 1996
{\it Rev. Mod. Phys.}{\bf 68} 1259

\bibitem{degennes99}
de Gennes P G 1999
{\it Rev. Mod. Phys.}{\bf 71} S374

\bibitem{kadanoff99}
Kadanoff L P 1999
{\it Rev. Mod. Phys.}{\bf 71} 435

\bibitem{wolf96}
Wolf D E, Schreckenberg M and Bachem A (ed) 1996
{\it Traffic and Granular Flow} (Singapore: World Scientific)


\bibitem{sugiyama08}
Sugiyama Y, Fukui M, Kikuchi M, Hasebe K, Nakayama A, Nishinari K, Tadaki S and Yukawa S 2008 
{\it New J. Phys. } {\bf 10} 033001 

\bibitem{wiesenfeld-moss95}
Wiesenfeld K and Moss F 1995 {\it Nature} {\bf 373}, 33

\bibitem{bulsara96}
Bulsara A R and Gammaitoni, L 1996 {\it Physics Today} {\bf 49}, 39

\bibitem{gammaitoni98}
Gammaitoni, L,  H\"{a}nggi P, Jung P and Marchesoni F 1998 {\it Rev. Mod. Phys.} {\bf 70}, 223

\bibitem{ohira-sawatari98}
Ohira T and Sawatari R 1998
{\it Phys. Rev. E} {\bf 58} 193

\bibitem{bonabeau99}
Bonabeau E, Dorigo M and Theraulaz G 1999 {\it Swarm Intelligence: From Natural to Artificial Systems} (Oxford: Oxford University Press)

\bibitem{glass88}
Glass L and Mackey M C 1988 {\it From Clocks to Chaos: The Rhythms of Life} (Princeton: Princeton University Press)

\bibitem{balachandran09}
Balachandran B, Kalmar-Nagy T and Gilsinn D E (ed) 2009 {\it Delay Differential Equations: Recent Advances and New Directions} (New York: Springer)

\bibitem{couzin05}
Couzin I D, Krause J, Franks N R and Levin S A 2005
{\it Nature} {\bf 433} 513

\bibitem{ohira00}
Ohira T and Yamane T 2000
{\it Phys. Rev. E} {\bf 61} 1247

\bibitem{ohira95}
Ohira T and Milton J G 1995
{\it Phys. Rev. E} {\bf 52} 3277

\bibitem{milton09a}
Milton J G, Townsend J L, King M A and Ohira T 2009
{\it Phil. Trans. Roy. Soc. A} {\bf 367} 1181


\bibitem{milton09b}
Milton J G, Cabrera J L, Ohira T, Tajima S, Tonosaki Y, Eurich C W and 
Campbell S A 2009
{\it Chaos} {\bf 19} 026110

\bibitem{mackey77}
Mackey M C and Glass L 1977 {\it Science} {\bf 197} 287

\bibitem{bando95}
Bando M, Hasebe K, Nakayama A, Shibata A and Sugiyama Y 1995 
{\it Phys. Rev. E. } {\bf 51} 1035 

\bibitem{nakanishi97}
Nakanishi K, Itoh K, Igarashi Y and Bando M 1997
{\it Phys. Rev. E. } {\bf 55} 6519 

\end{thebibliography}
\end{document}